# A SYSTEMATIC REVIEW OF UNCERTAINTIES IN SOFTWARE PROJECT MANAGEMENT

Marcelo Marinho[1, 2] , Suzana Sampaio[2], Telma Lima[3] and Hermano de Moura[1]

[1]Informatics Center (CIn), Federal University of Pernambuco (UFPE), Recife, PE, Brazil
[2]Statistics and Informatics Department (DEINFO), Federal Rural University of Pernambuco (UFRPE), Recife, PE, Brazil
[3]Administration Department (DADM), Federal Rural University of Pernambuco (UFRPE), Recife, PE, Brazil

## ABSTRACT

*It is no secret that many projects fail, regardless of the business sector, software projects are notoriously disaster victims, not necessarily because of technological failure, but more often due to their uncertainties. The threats identified by uncertainty in day-to-day of a project are real and immediate and the stakes in a project are often high. This paper presents a systematic review about software project management uncertainties. It helps to identify the difficulties and the actions that can minimize the uncertainties effects in the projects and how managers and teams can prepare themselves for the challenges of their projects scenario, with the aim of contributing to the improvement of project management in organizations as well as contributing to project success.*

## 1. INTRODUCTION

Project management has been discussed by executives and academics as one of the possibilities for organizations for integrating complex efforts and bureaucracy reduction. Managing projects effectively is introduced as a solution and as well as a major challenge for the business world. That is why both projects and project management have  a highly important role in society and have been used as scientific research objects.

Such subject is being applied in new industries, countries and spheres. Project management has become a core business process for a large number of companies both at strategic and operational level. We may call a project any activity that is considered significant and necessary and each major project can include a series of sub-projects.

Once software development involves a certain level of complexity and challenges, the techniques use, practices and project management tools have become common in software engineering. Nowadays it is very common for companies to deal with software development or service as a temporary project which needs to be planned, organized, conducted, monitored and controlled.

The Standish Group [1] reported that on average only 39% of projects are delivered on time, within budget and with the agreed requirements (therefore those projects perceived as successful). 43% are delivered late, and/or over budget and/or under certain conditions and finally, 18% are cancelled on delivery and never used. More than a decade later, very little seems to have changed.

Projects are seen as a strategic factors in organizations, even though, many projects still fail. One of the main reasons for such outcome is that project managers do not know how to deal with uncertainties [2]. Moreover, in project risk management literature, there is no common understanding as to what uncertainty is [3].

The term uncertainty may be understood broadly as "lack of certainty", which means information absence. Therefore, it covers not only probabilistic or indefinite results, but also an ambiguity and clarity lack concerning a large number of factors. Uncertainty is simply an ambiguity expression and project indeterminacy [4].

A systematic review discovers, evaluates and interprets all available research relating to the particular question. A systematic review aims to better understand and explore the topic, which is a planned and ideally repeatable way of synthesizing results from the existing body of scientific literature which has been prepared.

The authors developed a systematic review of related studies in the period from 1994 to 2013 guided by the research questions. The research aims to investigate: The relation of the uncertainty level associated with innovation projects; what the best practices to manage the uncertainties in software projects are; what are the sources of uncertainty perceived by studies; and what practices (techniques or strategies) are used for the problem nature  recognition and uncertainties containment in projects.

Following the introductory section, this paper is structured as follows: Section 2 is about Evidence-based-Software Engineering; Section 3presents the research method adopted for this study; Section 4 describes a data analysis extracted from the selected studies; in Section 5 the results for each research question  are presented and summarized  and  finally Section 6 contains the conclusion.

## 2. EVIDENCE-BASED SOFTWARE ENGINEERING

Evidence-based software engineering aims to provide means by which the best evidence from research can be integrated with practical experience and human values in the decision-making process considering the development and  software maintenance [7]. The essence of evidence-based paradigm is systematically collect and analyze all available data about a phenomenon for a more comprehensive and broader perspective than one can capture through a single study.

The paradigm based on evidence gained forces initially Medicine (Evidence-based Medicine/EBM), which aims to integrate the best research evidence with clinical experience and assessment of patients. Work kitchenham et al. [5] was the first to establish a parallel between Medicine and Engineering Software regarding to evidence-based approach. Kitchenham et al. [5] believe that software engineering can provide evidence-based mechanisms needed to help the professional to adopt appropriate technologies and avoid unsuitable ones, aiming the best practices and procedures. Some studies suggest that software engineering professionals (researchers) must consider the  use of evidence-based software engineering support to improve their decisions about which technologies to adopt [5],[6],[7],[8],[40],[41].

Software engineering based on evidence gathers and evaluates existing evidence in a technology through a five-step methodology. The steps 1, 2, 3 and 4 their evaluation (step 5) is done through a systematic review [40]:

- Transforming the problem or need for information on a research question;
- Search the literature for the better available evidence to answer questions;
- Critically evaluate the evidence about its validity, impact and applicability;
- Integrate the evaluated evidence with practical experiences, values and clients circumstances  to make decisions;
- Evaluate the steps of 1 to 4 performance and search for ways to improve them.

## 2.1. Systematic Literature Review

Systematic literature reviews evaluate evidence in a systematic and transparent way. In a traditional literature review, the research strategy and results evaluation criteria are usually hidden from the reader, which means that the revision may perfectly be done in an unstructured way, ad hoc and evidence that do not support the researcher's preferred hypothesis might be ignored. However, in a systematic literature review the research strategy and the evaluation criteria are explicit and all relevant evidence are included in the evaluation [5],[6],[7].

A systematic literature review "is a way of evaluating and interpreting all available research relating to a particular research question, topic area, or phenomenon of interest" [42]. Travassos [43] believes systematic reviews "provide the means to perform comprehensive literature review and not biased, giving their results have scientific value". Kitchenham [5] adds that systematic reviews aim to present a fair assessment of a research topic using a reliable, accurate and auditable methodology.

Kitchenham [5] e Travassos [43] present some of the reasons for conducting a systematic review:

- Summarize existing evidence about a phenomenon;
- Identify gaps in current research;
- Provide a framework to position new research; and
- Support the generation of new hypotheses.

According to Kitchenham [5], his early studies were based on the guidelines used in the medical field, however it is important to recognize that software engineering research have large differences from medical research, so an approach that incorporates guidelines of social science researchers  was added in their studies and more current recommendations. Kitchenham [5] summarizes the steps of a systematic review in three main phases: Planning the review, Conducting the revision and Presenting the revision. These steps are described below.

## 2.2. Planning a Systematic Review

As in any scientific endeavor, a systematic review of the literature needs a detailed protocol describing the process and the methods to be applied. The most important activity during the planning phase is the formulation  of research questions to be answered as all the other aspects of the review process depend on them [40]. To Kitchenham \cite{kitchenham2004procedures}, before undertaking a systematic review researchers must ensure that it is necessary and the protocol should be able to answer some questions:
- What are the objectives of this review?
- What sources were searched to identify primary studies? Were there any restrictions?
- What were the criteria for inclusion / exclusion and how they are applied?

- What criteria were used to evaluate the quality of the primary studies?
- How were the quality criteria applied?
- How was the data extracted from primary studies?
- How was the data synthesized?
- What were the differences between studies investigated?
- Because the data were combined?

Through these and other questions, the researcher plans and documents all necessary information to carry out the systematic review.

## 2.3. Planning a Systematic Review

Once the protocol has been defined and validated the review can begin. The primary studies selection, or else, the execution of the selection process defined in the protocol for the pursuit of studies and subsequently data extraction and evaluation are part of the implementation phase of the systematic review execution. For the studies selection, inclusion and exclusion criteria are used. The information extraction and evaluation are conducted through forms and may be supported by a software tool. The steps, according to Travassos [43], summarized for the review implementation are:

- Searches in the defined sources: the process should be transparent, repeatable and documented, as well as the changes that occur in the process;
- Primary studies selection with the inclusion and exclusion criteria defined;
- Data extraction from general information studies to answers to the research questions. Forms are a good way to record all the necessary data and the use of a computational tool can support the data extraction and recording and subsequent analysis;
- Assessing the studies quality is important to balance the importance of different studies, reduce bias (tendency to produce ``biased results'' that systematically separates from true results), maximize internal and external validity and guide recommendations for future research;
- Data synthesis is performed according to the research questions, through tables to highlight the similarities and differences between studies. If quantitative data are available, making a meta-analysis can be considered.

## 2.4. Presenting the results

The last step of a systematic review consists on writing a review report and its evaluation, according to the synthesis, and data analysis. Lately, the results are consistently presented with tabulated information and the research question, highlighting similarities and differences between the results, or else, highlighting the possible data combination and analyzes.

## 3. RESEARCH METHOD

This section describes the research method according to Kitchenham methodological guideline for systematic reviews [8]. A systematic review protocol was written to describe the plan for the review. Details on the course of these steps are described in the following sub-sections.

## 3.2. Research questions

These are research questions which guided the systematic review:

- **RQ 1:** What is the relation between uncertainty and innovative projects?
- **RQ 2:** What are the sources of uncertainty perceived?
- **RQ 3:** How is it possible to reduce the uncertainty level in software projects?

- **RQ 4:** What practices (techniques or strategies) can help reduce the uncertainties in project management software?

## 3.2. Search environment

A directory in the cloud was created before starting the research. A free web store service was used by the researchers to store all artifacts used. This enabled a total standardization and control, so all of them could access the material as if they were in the same environment, thought being remote.

Furthermore, some datasheets were developed to be used in all phases. Those datasheets facilitated the data organization in many aspects; for example, a standard to enumerate searched publications, filters to extract objective information among others.

## 3.3. Generation of Search Strategy

Aiming to identify and recover the smallest possible publications super group which meet the systematic review eligibility criteria, it incorporates a search strategy for a research. The eligibility criteria are conditions to determine if primary studies are about the systematic review research questions. The search results are transformed into in a sequential publication list of the chosen engines. Each resource has a different community, interests, language, examining issues and search syntaxes as well. Therefore, different resources might require different search strings.

Next, we conducted initial studies for all phases of the major study, called "pilot studies". These were performed to align a phase-to-phase understanding among researchers, all search engines mechanisms test and the adjust of some search terms. The study only proceed when the two researchers agreed with the pilot results.

The resources used to searches are: IEEEXplore Digital Library (htttt://ieeexplore.ieee.org/); ACM Digital Library (http://portal.acm.org); Elsevier ScienceDirect (www.sciencedirect.com); Springer Link (http://link.springer.com).

Other sources were initially considered as potential for searches:Google, Google Scholar, Wiley InterScience, InspecDirect, Scopus and Scirus. However, these were later excluded from the final list of sources for some of the following reasons:

- Some because they were not present in significant or systematic reviews or not having been recommended by experts;
- Some for not allowing the viewing or downloading of the works without payment or license that the institution responsible for the work did not have;
- Some because they were already indexed by some of the sources already listed in the search.

To search all results from sources, the researchers grouped to search publications. The sources (engines) were divided among them. Each researcher was responsible to find results in their engine and, finally the papers were catalogued. Then, when the search was performed, there were identified 3,044 publications. The searches results were extracted in Bibtex files to merge in a datasheet developed to consolidate the results of all engines. After excluding duplicated results from the datasheet, we found 2,933 articles to start the first phase.

## 3.4. Paper Selection

The idealized selection process was done in two parts: an initial document selection of the research results that could reasonably satisfy the selection criteria based on a title and the articles abstract reading, followed by a final selection of the initially selected papers list that meet the pre-established criteria, based on the introduction and conclusion reading of the papers. To reduce potential bias, the process was conducted in pairs, in which both researchers worked individually on the inclusion or exclusion of the paper and then a comparison of spreadsheets was done. The possible divergence was discussed and finally a consensus was reached. If there was not a consensus, a third researcher should be consulted. In case doubts still remained, the work would be inserted in the list.

In the pilot study performed before the first phase beginning, the first ten results from all engines were catalogued and the group read the titles and the abstracts and discussed about them to calibrate comprehension. Other pilot study was performed having more five publications done, because the researches were not ready to continue after the first pilot. After a reliability agreement, the first phase initiated. Each researcher read the publications` titles and abstracts to select or exclude them. They discussed about the results to gather them together according to a new datasheet agreement. Out of the initial selection of 2,933 papers, 111 articles were selected to the second phase.

After the first and before second phase, a new pilot study was done. Then, we selected a single article to be read by researchers aiming a consensus for both. In this phase, the introduction and the conclusion should be read. Similarly to the first phase, each researcher read the articles individually and later discussed its results. After the phase two selection, the researches eliminated 88 and selected 23 papers out of 111 to be read for data extraction phase.

## 3.5. Study Quality Assessment

In the data extraction phase, each publication methodological quality of was assessed. Three factors were assessed as follows, and each were marked **yes** or **no**: Does the publication mention the possibility of selection, publication, or experimenter bias?; Does the publication mention possible threats to internal validity? and Does the publication mention possible threats to external validity?

The quality assessment was made only on the basis of whether the publication explicitly mentioned those issues. We did not make judgement about whether the publication had a "good" treatment of the issues. The results of the study quality assessment were not used to limit the selection of publications.

## 3.6. Data Extraction

Beforehand, a new pilot was done to calibrate the stage. We selected two relevant articles and compared the extraction data performed so far with ours. Thus, a pitot was carried out with an article found with one of the 23 selected works. In the data extraction phase, researchers had to read the papers selected for extracting information according to the datasheet model.

We had selected 23 works, but during the extraction phase we identified 2 articles that showed no relevant citations or possible reasons to be extracted, thus, there were 21 articles. For each publication there was extracted information.

From each study, a list of shares was extracted, where each share described answered a research question. Or else, each simple sentence that answered one or more research question was considered a quota. We had a total of 165 quotas extracted from 21 studies. Finally, these shares were recorded on a datasheet.

### 3.7. Data Synthesis

The two researchers worked on the synthesis work to generate combinations of quotas with answers of the research questions.

There was a good level of inter-rater agreement, differences in opinion were discussed in a meeting, and it was easily solved without the need of involving a third arbitrating researcher, as planned.

## 4. SYSTEMATIC REVIEW RESULTS

This section describes the analysis of the data extracted from our selected studies.

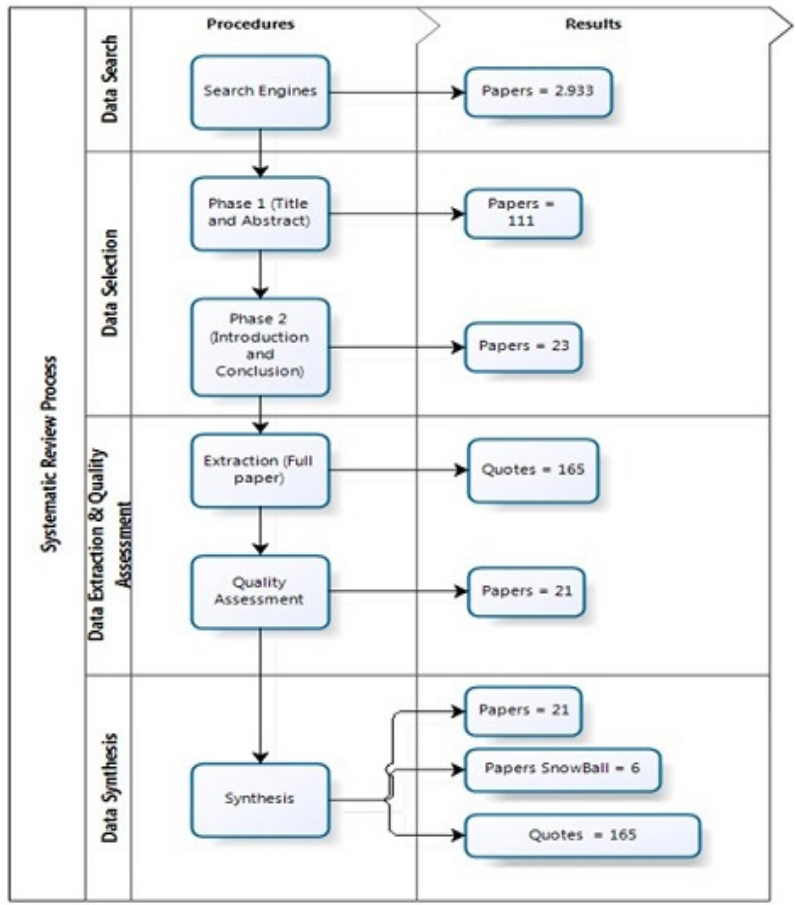

Figure 1. Results obtained on each stage at systematic review process.

## 4.1. Data Search

In the Data Search phase the searches were conducted in four sources. The Figure 1 shows the results obtained on each stage at systematic review process.

## 4.2. Data Selection

The Data Selection was divided in two phases: Phase 1: Title and Abstract analyses; and, Phase 2: Introduction and the Conclusion analyses.

In Phase 1, after checking the titles and abstracts, 111 articles were selected for the next phase. A total of 2,822 articles were eliminated. Among the criteria used, we may highlight: "0 - Not applicable" (this is a research non related with management or uncertainties), with 52%; "Outside the uncertainties in project management area", with 38% and "2 -It is about this risks in projects" with 10%.

In Phase 2, after reading the introduction and conclusion, just 23 papers were selected for the extraction phase. A total of 88 articles were eliminated. Among the criteria used we highlight: "1- Outside the uncertainties in project management area" with 59% , "2 - It is about this risks in projects" with 34%, followed by "0 - Not applicable", with 7%.

Table 1. List of engines and its absolute contributions.

| Engine | Automatic | Selection 1 | Selection 2 | Extraction |
|---|---|---|---|---|
| ACM | 548 | 10 | 2 | 2 |
| IEEE | 722 | 63 | 15 | 13 |
| ScienceDirect | 569 | 11 | 4 | 4 |
| Springerlink | 1094 | 27 | 2 | 2 |
| Total | 2933 | 111 | 23 | 21 |

## 4.3. Data Extraction and Quality Assessment

In the Data Extraction and Quality Assessment each research read a full paper for the quotes extraction, at the same time they did the paper quality assessment. At the quality assessment phase three factors were assessed as follows, and were each researcher marked YES or NO: i) Does the publication discuss the possibility of selection, experimenter, or publication bias?; ii) Does the publication discuss threats to internal validity?; and, iii) Does the publication discuss threats to external validity? The quality assessment was made only on the basis of whether the publication itself mentions with these issues – we did not assess whether the publications had a treatment of these issues.

Out of 23 works selected in the previous stage, the researchers worked with 21 articles. Two were eliminated for not answering the research questions. Additional studies were identified by search technique snowball, or else, covering the studies references already found. This technique allowed us to identify high-quality works that were not found by the automatic search. 6 works that contributed to the discussion were added to be held in Section 5. 6 works out of 4 are books and 2 are journals that are not indexed by engines the selected research.

It is worth pointing out that some of the non-selected studies for extraction phase were considered so, because they confused risk with uncertainty. Note that although the risk and uncertainty are related, they are not the same thing. The uncertainty is the unknown, whereas risk is what can go wrong. Clearly, much of the risk of the project depends on the uncertainty, but there are other factors that contribute to the risk of the project, including deadlines, lack of resources and inadequate skills. Several of the articles report "uncertainties in the estimates" when in fact they are trying to estimate risks in the project process.

## 4.4. Data Synthesis

Considering the studies evaluated in the extraction and adding the works found by snowball in the Synthesis phase 165 quotas were analyzed, in which 18  answer the first research question; 44 the second; 30 the third and 73 the last.

Figure 2 represents the distribution of works by continent. The geographical distribution of the uncertainties related to studies in project management was as follows: The United States was ahead with 13 articles; England, with 3; Brazil, China and Singapore with 2; and Australia, Scotland, Philadelphia, Israel and Pakistan, with 1.

We note that in the last 10 years it has been published 20 out of 27 of the papers included in the study. It demonstrates and confirms that research on uncertainties in project management have been growing since the last decade. Figure 3 illustrates the distribution of studies identified by selection process throughout the years.

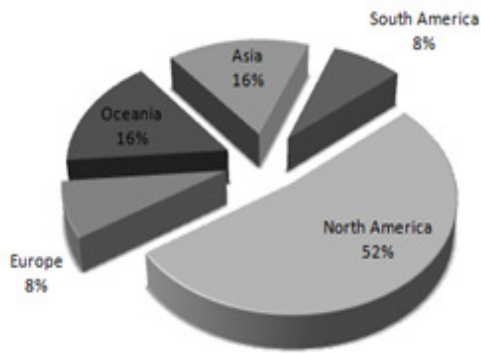
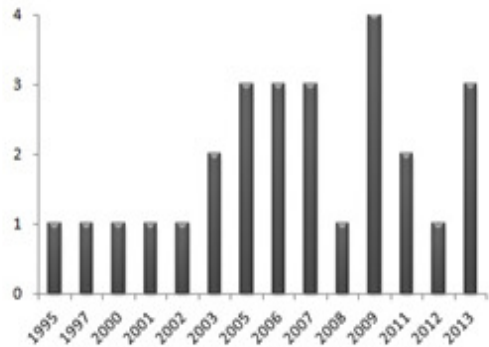

Figure 2. Distribution by continent.          Figure 3. Distribution by years.

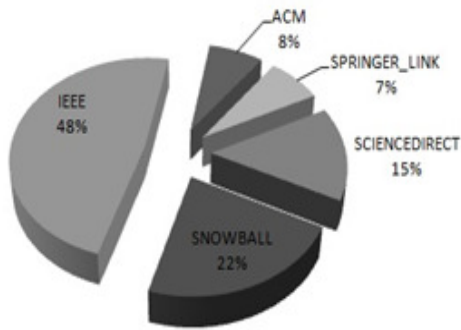
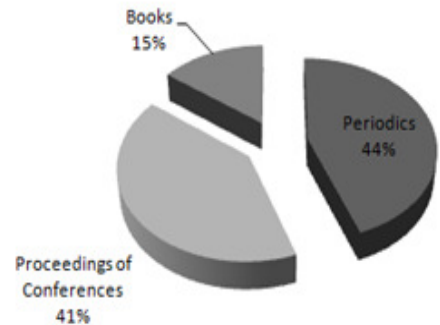

Figure 4. Distribution by engines.      Figure 5. Distribution by type of publication.

Figure 4 shows the distribution of documents found by search engines and Figure 5 shows the distribution of jobs by type of publication, showing that most of the studies, 44%, were published in journals, following a 41% of the annals of events (Conferences, Workshops and symposia) and 15% in books. Figure 6 shows the papers`methodologies adopted. Table 2 displays a list with the number of studies returned by event. It can be noticed that most of the studies were published in some editions of the PICMET annals *Portland International Conference on Management of Engineering and Technology*, followed by the IJPM periodicals *International journal of project management* and the third, JPIM *Journal of Product Innovation Management*.

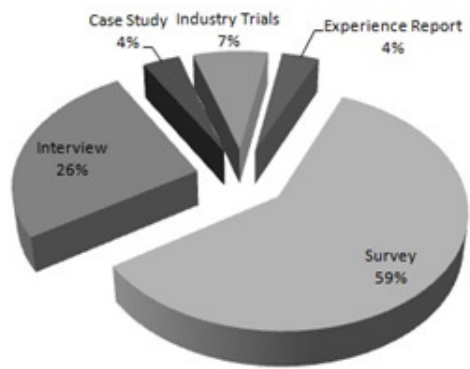

Figure 6. Distribution by methodologies.

## 5. FINDINGS

All evidence is properly referenced by 27 studies [9], [10], [11], [12], [13], [14], [15], [16], [17],[18], [19], [20], [21], [22], [2], [23], [3], [24], [25], [26], [27],[28], [29], [30], [31], [32], [33]. In this section, the results for each research question are presented. In Section 5.1 the evidence that the uncertainties become larger and larger as the project is more innovative are presented. In Section 5.2 sources of uncertainties are perceived in the studies. In Section 5.3 the evidences are presented about the possibility of reducing the uncertainties in software projects. In Section 5.4 evidences on how the techniques or strategies that favor uncertainties reduction in project management are presented.

Table 2.  List with the number of studies returned by event.

| Published in | Amount |
|---|---|
| Portland International Conference on Management of Engineering & Technology | 5 |
| International journal of project management | 4 |
| Books | 4 |
| Journal of Product Innovation Management | 3 |
| Engineering Management Conference | 2 |
| International Conference onSoftware Engineering -Workshop on Cooperative and Human Aspects on Software Engineering | 1 |
| IEEE International Conference on Systems, Man, and Cybernetics (SMC) | 1 |
| IEEE Software | 1 |
| IEEE Transactions on Engineering Management | 1 |
| IEEE Transactions on Systems, Man and Cybernetics | 1 |
| International Conference on System Sciences | 1 |
| MIT Sloan Management Review | 1 |
| Small Business Economics | 1 |
| SIGPLAN conference companion on object oriented programming systems languages and applications | 1 |

## 5.1. What is the relation between uncertainty and innovative projects?

This question sought to investigate the relationship of the uncertainties with innovative projects and aspects of how we manage these projects. One of the findings is that the evidence report that innovative projects are more likely the occurrence of uncertainties. The following evidence of the studies is reported.

Innovation and projects aimed at the innovation development, being them a new product, process or service, should be on the executive diary, along with the understanding of the business environment changes and the action plan needed to respond to these changes or influence them [31].

Various perspectives emerge from the literature to explain why companies have difficulty in managing the various sources of uncertainty associated with converting innovations in innovative companies. Understanding the innovative project characteristics and the uncertainty nature that permeates them is critical for developing appropriate management practices [32].

Each project has different characteristics, so the uncertainty level is varied. High-risk projects, typically innovation projects, which normally pursue ambitious goals and high failure probability, will have a high uncertainty capacity [4]. Low uncertainty projects typically create limited opportunities, whereas a project of high uncertainty will be evaluated primarily on their commercial effect and long-term, rather than measures of time and budget [34].

The largest uncertainty occurs in projects in which the company has no previous experience or existing project on which to build. On the contrary, the projects in which only a small amount from the existing project is changed, the uncertainty to the development teams is reduced [14].

Where uncertainty about future events is high, tolerance of uncertainty may be particularly necessary [19]. Management practices that are effective in established  companies are often ineffective and even destructive when applied to innovative projects because of high levels of uncertainty inherent in these projects [32].

Flexibility is required on projects where goals are unclear or open to negotiation, the strategy is emerging and the project is highly subject to external influences. Keeping options open, and adopt a flexible and robust management approach can be much more effective than prematurely freeze projects plans relying on conventional control mechanisms to measure performance [19].

Strategies, techniques, best practices are important to manage or contain the uncertainties in innovative projects. The recognition of the uncertainty sources  in projects can be a determining factor in the success of the project. You need to clarify what can be done, decide what should be done, and to ensure that management is carried out based on the uncertainty level observed in the project.

## 5.2. What are the sources of uncertainty perceived?

This question sought to investigate what the sources of perceived uncertainty in software projects are. From the 27 studies analyzed, 44 quotes was found. In the studies surveyed there is not a single label to the sources, it is common in some articles to speak from a particular source that is a single uncertainty, but classified differently, for example: market source [21], [17], [25], [22], [31], [32], external source [33] and  novel source [11],[2] represent a single source of uncertainty. For doing so, we try to group the uncertainty and create a unique classification for these sources. Among those quotas found (they could cite more than one source, if they are in the same sentence), we find: 16 references to market uncertainty; 15 references to technological uncertainty; 14 for environment uncertainty and 9 for socio-human uncertainty. The following describes the sources of uncertainties grouped by research:

### 5.2.1. Market

Market uncertainty indicates how news products are to the market, to consumers and to potential users. It represents the extent to which buyers and users are familiar with this type of product, its benefits and how they can use it. The level of market uncertainty indicates the external uncertainty, and also reflects the uncertainty of the project goal.  It also indicates the easiness of knowing what to do or what to build and how to introduce consumers to the product. The market uncertainty comprehends client, suppliers, partners and current market situation [11],[13], [21], [17], [25], [22], [2], [29], [31], [32], [33].

### 5.2.2. Technological

Technological uncertainty depends on the extent to which the project uses new technology or mature technology. The level of project technological uncertainty is not universal, but subjective, and depends not only on what technology exists, but also what is accessible to the organization. It is therefore a measure of the amount of existing new technology compared to mature technology available for use in the project. The technological uncertainty causes, among other things, an impact in the project, in the communication, in  the freezing time of the plan and the number of planning cycles. It can also affect the technical expertise that the project manager and their team members need to have [10], [11], [13], [16], [17], [21], [22],[25], [31], [32].

### 5.2.3. Environment

This source indicates the degree of uncertainty of the external and internal organizational environment. Organizational theories emphasize that organizations must adapt to their environment if they are to remain viable. Environmental uncertainty can arise from the actions of different organizations (suppliers, competitors, consumers, government, shareholders, etc.) and this may affect the project. Doubts about the probability or nature of changes in the environment (socio-cultural trends, demographic changes) can lead to a number of environmental uncertainties [11], [16], [21], [29], [31], [32].

### 5.2.4. Socio-Human

Although modern organizations have technological tools that can meet most needs and structural deficiencies, it alone is not enough to ensure the acquisition of individual and group knowledge. This is due to cognitive factors intrinsically related to how people perceive, learn, remember and think about information. Projects can be a unique way of helping organizational processes to change, innovate and adapt to competitive market reality.

Human relations are often viewed as fuzzy in the management process. When misunderstood, they can lead to conflicts that can threaten project development, especially in technological innovation projects that present a high degree of complexity and uncertainty. This kind of challenge requires creativity and flexibility in project teams [31].

The socio-human source considers the relationships between people within an organization. It is necessary to consider religious issues, politics, different values, personal experiences and cultural training. Any of the mentioned factors can affect project performance and results. Project managers need to deal with social differences and avail themselves of each team member's particularity and their potential in order to assist in the project success everyone [26], [28], [30], [31].

### 5.3. How can you reduce the uncertainty level in software projects?

This question aimed to investigate the possibility of reducing the uncertainty level in software projects. From 27 studies analyzed, 30 quotes were found and classified. There are 5 ways to manage uncertainties in projects identified by the research. They are: 9 approach adopting techniques and strategies to facilitate the uncertainties reduction; 8 address adapting management style to the projects type; 6 approach dealing with uncertainty when they happen; 5 approach understanding the uncertainty sources to better manage each type of project; 2 address identifying uncertainties in order to turn them into risks.

### 5.3.1. Adopting techniques and strategies to facilitate the uncertainty reduction

It is tempting to wish to eliminate all uncertainty, but the high levels of necessary resources even to get close to project goal are, in the most exceptional cases, unjustified. In fact, great efforts in the uncertainty sources eradication often divert attention from the real goals. Eradication is rarely the answer, it is more feasible to manage uncertainty within acceptable levels. This leads to another guiding principle for the uncertainty management: The objective is the uncertainty containment, not its elimination [13],[18], [22], [2], [23], [3], [29], [33].

Although there are no easy answers or fast solutions, we may say that uncertainty can be ``tamed'', part of the answer lies in recognizing the nature of the problem and select the right technique (or strategy) to work. Like any good craftsman, the project manager must have of a set of comprehensive tools for managing uncertainty and - equally important - a good knowledge of the capabilities and limitations of those tools [29]. For different types of problems, the manager and the team should have strategies, mindset and different paradigms [18].

### 5.3.2. Adapting management style to the type of projects

Many projects with all the ingredients for success still fail. The reason for this is that executives, project managers and the project teams are not used to assessing and analyzing uncertainties, so they fail to adapt their management style to the situation. Shenhar et al [34] affirms that "the same size does not fit all", meaning that every project is unique, and that one must understand how they differ and take the proper actions according to the particular needs of the organization and the project.

The authors described the importance of evaluating and analyzing a project's uncertainties and complexities and so, adapting their management style to situation [31]. It is also addressed in some studies that projects fail because managers applied the wrong management style to the project [18]. It is important to point out that project managers cannot predict the future, but can perceive the uncertainty degree in their projects and choose an appropriate management style to use [10],[13].

### 5.3.3. Dealing with uncertainty when they happen

Some uncertainty types cannot simply be solved  through an analytical approach. Such as:   a number of events, random combinations, may contribute to an unexpected result. Pharmaceutical companies have struggled with this problem. Despite extensive testing programme, there is always the risk of an unlikely combination of external factors (other drugs usually administered by the patient) to react causing harmful side effects [21], [22], [2], [3].

Project managers can try to contain uncertainty at its font but they can hardly ever have a hundred percent  of success. Therefore, a project needs strength and should be able to rapidly detect and respond to unexpected events. A project manager must decide how best to cope with unexpected results  [29],[31].

### 5.3.4. Understand the sources of uncertainty to better manage each type of project

Uncertainty can arise from deficiencies in various sources, such as contextual information about the project, our understanding of underlying processes, past events explanations and speed (or time) change. We may ask where these factors happen within a project and what aspects of a project plan are particularly vulnerable to each uncertainty type. To answer those questions, first it is interesting to see the elements that make a typical project. Then, we need to examine what happens when the scale and complexity of the model increases. Thus, it is possible to choose styles and strategies to manage the project properly [11], [13], [22], [15], [2], [23], [31].

### 5.3.5. Identify uncertainties in order to turn it into risk

Strategies can be used to contain the uncertainties. These strategies can help you learn more about the nature of uncertainty, for example, through the formulation of the problem that it represents or the modeling of future scenarios to prepare for them. Once an uncertainty is revealed, analytical techniques, such as risk management can be used in project management [11], [23].

## 5.4. What practices (techniques or strategies) can help reduce the uncertainties in project management software?

This question sought to identify practices to support the software projects management that help reduce the uncertainties in innovative projects. Out of 73 quotas extracted for that matter, 18 practices for managing projects focusing on reducing uncertainties were found. These practices are described below together with their references of studies that support each of them. We presented an evidence before the explanation.

### 5.4.1. Identifying the project type to adopt appropriate management

*"It is necessary to understand in what way projects are different from each other in order to suit the right situation to the particular project."[31]*
To reduce the failure probability of a project it is important to characterize it properly, so knowing in advance if there is a related uncertainty to their goals and solutions adopting a management model that fits the project type. [9], [10],[18], [19], [2], [23], [3], [29], [27], [28], [31], [33].

### 5.4.2. Managing the stakeholders' expectations so that they flexibly accept changes

*"With uncontrolled uncertainty, a lot of time and effort must go into managing relationships with stakeholders and getting them to accept and respond to unplanned changes." [18]*
*"A lot of time and effort must go into managing relationships with stakeholders and getting them to accept unplanned changes."[13]*
Innovative projects can create high expectations for clients. You need to manage them, keeping clients informed and aware of the project uncertainties, as well as creating a bond of trust between project members and clients [13], [18], [19], [21], [23], [3], [24], [27], [33].

### 5.4.3. Ability to formulate qualitative measures of success

*"Ability to formulate qualitative success measures for projects is another tool that should be added to the project management armoury to assist in managing softer projects."[19]*
Projects with low uncertainty can often be assessed through quantitative success measurement, such as time and cost, and tangible performance measurement related to their tangible final deliverables. Projects with high levels of uncertainty require different forms of performance evaluation that recognize the validity of different perspectives and worldviews [19], [2].

### 5.4.4. Identifying early warning signs to manage the uncertainties

*"to identify unforeseeable uncertainties is through the perception of early signs."[33]*
The objective is to identify the cause of problems and the solutions for each early signal [22], [3], [28], [33].

### 5.4.5. Sensemaking

*"understanding and sensemaking affect strategic decisions, and consequently, performance of the firm."[3]*
Sensemaking is the process by which organizations and individuals work out uncertainties, ambiguities, changes and problem situations generating inventions and new situations that end up in actions that lead to problem solution and environmental stability. [22], [3], [28], [33].

### 5.4.6. Management flexibility and ability to respond to changes

*"Each manager tends to build his or her own's mental model of the project and its challenges, albeit based on his or her experience and professional judgement" [12]*
The flexibility and ability to communicate the changes is fundamental. Complex and uncertain projects changes require greater flexibility and reflection. The project manager and the team performance should change as the profile and the uncertainty evolve. [12], [13], [17], [20], [21], [2], [3], [31].

### 5.4.7. Managerial ability to perceive uncertainty and deal with them

*"A preliminary observation of the areas of uncertainty can help project managers and teams to conduct a successful project." [31]*
The ability the project manager should have to take reasonable decisions to ensure there is necessary support to get everyone involved in the project; they must have personal ability, such as intuition and trial to perceive uncertainties [18], [12], [13], [21], [3], [26], [31].

### 5.4.8. Team willing to learn and develop new ideas in order to generate knowledge

*"When enough new information arises, they must be willing to learn and then formulate new solutions." [13]*
Crisis management and continuous observation of threats and/or opportunities must be controlled by the team . When new information arises, everyone should be willing to learn and then formulate new solutions [12], [13], [18], [3], [24], [31], [33].

### 5.4.9. The creation of flexible contracts

*"The manager's job is to anticipate and soften resistance by creating flexible contracts and keeping stakeholders well informed" [13]*
Creating flexible contracts for innovative projects help mitigating resistance to changes necessary for the project. In order to keep project stakeholders well informed, it is obviously important to have a flexible contract [13],[33].

### 5.4.10. Building trust between team, management and customer

*"The relationship is characterized by trust."[13]*
The relationship between clients, managers and teams is characterized by trust. Once it is conquered, helps facilitate the strategies change during the project [13],[21],[28].

### 5.4.11. Verify information outside environment of the project

*"Judging the source and relevance of information that comes from the outer project environment and, thus, represent contextual uncertainty is an intuitive process rather than a rational one, since the rational processes are isolated from the surrounding world." [3]*
Relevant actions  to organizations or groups within the organization (suppliers, competitors, consumers, government, shareholders, etc) can affect the product,  as well as doubts about the likelihood or nature of changes in the environment general condition (socio-cultural trends, demographic changes) [3],[13], [2], [28], [31], [32].

### 5.4.12. Understanding the sources of uncertainties

*"it is necessary to understand the areas of uncertainty in a project to be able to contribute to its success." [31]*
Project management can be conducted focusing on solving the project uncertainties; for doing so, it is necessary to understand where the uncertainties of projects can arise, ie, what the possible sources of uncertainty are. Understanding the sources we may be able to make the necessary changes as the project progresses [14], [2], [31], [32].

### 5.4.13. Project Managers must incorporate the investigation of uncertainties in projects

*"The use of uncertainty management within project management can be a determining factor in project success." [31]*
Ongoing uncertainties investigation is important for projects members act in a proactive way and for the organization benefit strategically. The articles show that managerial knowledge aligned with research uncertainties may contribute to transformation of uncertainties in risk [20], [2], [31], [32], [33].

### 5.4.14. Learning method

*"learning strategies: give the project manager, and the organization as a whole, the ability to improve and benefit from experience over time." [4]*
This is about experimentation and improvisation. The more we experience knowledge of a particular subject the more we reduce uncertainty. Improvisation itself can be a differentiator for innovative projects to deviate from uncertainties, leading to new goals [4],[18],[28],[33].

### 5.4.15. Creativity techniques

*"Encourage diversity and stimulate the creativity of empowered team members." [18]*
Some articles suggest techniques to obtain knowledge such as: brainstorming, feasibility study, market research [18],[28],[33].

### 5.4.16. Managers should facilitate communication within the organization

*"Managers need to participate in and facilitate freeflowing communication and conversations which are absolutely critical to success." [18]*
Some articles suggest that innovative projects with small teams and located in the same environment easily pass on the information received [13], [17], [18], [3].

### 5.4.17. Managers should facilitate self-organization and the team adaptability

*"The managers should facilitate evolutionary, double-loop learning, and facilitate self-organization, innovation, and adaptability" [18]*
They need to encourage diversity of thought and interaction, breaking organizational and hierarchical structures. The team need to adapt to changes and to interact getting feedback with market and technology providers constantly. The managerial focus should be on group dynamics to keep large project objectives in mind, instead of labor control [18], [28].
### 5.4.18. The Collaborative Work

*"The development team is collocated in a collaborative workspace" [24]*

The democratic management style is best appropriated; very tight control might lead an innovative project clutter and the project vision become an illusion. Collaborative working is essential in projects with many uncertainties [18], [2], [24].

# 6. CONCLUSIONS

This paper presented a systematic review of uncertainties in project management software. Various approaches to project management does not consider the impact that uncertainty has on project management. The threats posed by uncertainty are real and immediate and the stakes in a project are often important. The project manager faces a dilemma: decisions must be made now about future situations that are inherently uncertain. The use of uncertainty management in project management can be a determining factor in the success of the project.

This research aimed to contribute with to the uncertainties management in project management software, contribute to software engineering based on evidence. The paper presented a detailed process for conducting a systematic review of literature that can be followed by other researchers for other works.

Finally, the results of this research contribute to the project management software in two ways. Firstly, the results of the systematic review provide to the academic community a better understanding of the challenges of dealing with uncertainties in project management and, thus, show gaps in the area that may be good opportunities for future research. Secondly, how to deal with the uncertainties, the techniques and strategies presented and sources of uncertainties that can support practitioners and researchers in identifying relevant challenges and developing solutions for projects, making use of the best practices that have been tested by other primary studies in experimental and industrial environments.

# ACKNOWLEDGEMENTS

Authors would like to thank CAPES for supporting the development of this work as well as the reviewers.

**Authors**

**Marcelo Marinho** is currently assistant professor at Department of Informatics (DEINFO) at Federal Rural University of Pernambuco in Brazil (UFRPE). Conducts doctoral research at Informatics Center at Federal University of Pernambuco (CIn - UFPE). in Software Project Management area, where he investigates Uncertainty Management in Software Projects. He holds a Masters in Computer Science (CAPES Concept 6) at University Federal of Pernambuco (UFPE) and graduated in Computer Science. He had been worked in the Software field for 10 years in which 6 with software project management. Besides, he has interest in the following lines of research: project management, software project management, strategic planning of information systems, software development process, IT management, software process and agile project.

**Suzana Sampaio** has 17 years of experience on the software development industry, working for the last 9 years as a consultant in process improvement, agile methodology and project management. She is currently a PhD candidate at Informatics' Center at Federal University of Pernambuco (CIn-UFPE). Her research focuses on Software Project Management where she investigates software project actuality. Also works as an assistant professor at Department of Informatics (DEINFO) at Federal Rural University of Pernambuco in Brazil (UFRPE). She holds a Masters in Computer Science (CAPES Concept 6) at University Federal of Pernambuco (UFPE), on Software Development Productivity (2010) and graduated in Computer Science (2000). She is an ISO 9001:2008, ISO 29110 Lead Auditor, MR-MPS-SW and CMMI implementer and appraiser. Besides, she has interest in the following lines of research: project management, agile project management, process improvement and quality assurance.

**Telma Lima** is lecturer at the Federal Rural University of Pernambuco (UFRPE), Brazil. She holds a Doctorate in Production Engineering from Federal University of Pernambuco (UFPE), Master in Production Engineering from UFPE and Bachelor in Mathematics from UFRPE, Brazil. She is currently acting as lecturer, teaching Decision Support Systems and Knowledge Management, and as researcher. Her research interests include negotiation, partnerships, innovation in new technologies, entrepreneurship, startups and  knowledge management

**Hermano Moura** is an Associate Professor and Prorector for Planning and Finance, Federal University of Pernambuco (Recife, Brazil). PhD in Computing Science (1993), University of Glasgow, Scotland. MSc in Computing Science (1989), Federal University of Pernambuco, Brazil. Electronic Engineer degree (1982) from Federal University of Pernambuco, Brazil. PMP – Project Management Professional (2003) by Project Management Institute. Research areas of interest includes: management of software projects; software process improvement; strategic planning of information systems. Acted as a consultant on various projects involving definition and implementation of processes for project management and software development.